\RequirePackage{fix-cm}
\documentclass[natbib,smallextended]{svjour3}       
\smartqed  
\usepackage{graphicx}
\usepackage{natbib}
 \journalname{my journal}
\begin{document}

\title{Future Simulations of Tidal Disruption Events}



\author{Julian H. Krolik \and Philip J. Armitage \and Yanfei Jiang \and Giuseppe Lodato }


\institute{J.H. Krolik \at
              Department of Physics and Astronomy, Johns Hopkins University, Baltimore MD 21218 USA \\
              \email{jhk@jhu.edu}           
           \and
           P.J. Armitage \at
              Department of Physics and Astronomy, Stony Brook University, NY 11794 USA\\
              \email{philip.armitage@stonybrook.edu}
           \and
            Y. Jiang \at
             Kavli Institute for Theoretical Physics, University of California at Santa Barbara, Santa Barbara CA 93106 USA\\
           \email{yanfei@kitp.ucsb.edu}
            \and
             G. Lodato\at
             Dipartimento di Fisica, Universita di Milano, Via Celoria 16, 20133 Milano Italy\\
             \email{giuseppe.lodato@unimi.it}
}

\date{Received: date / Accepted: date}

\maketitle

\begin{abstract}
Tidal disruption events involve numerous physical processes (fluid dynamics, magnetohydrodynamics, radiation transport, self-gravity, general relativistic dynamics) in highly nonlinear ways, and, because TDEs are transients by definition, frequently in non-equilibrium states.  For these reasons, numerical solution of the relevant equations can be an essential tool for studying these events.  In this chapter, we present a summary of the key problems of the field for which simulations offer the greatest promise and identify the capabilities required to make progress on them.  We then discuss what has been---and what cannot be---done with existing numerical methods.  We close with an overview of what methods now under development may do to expand our ability to understand these events.
\end{abstract}

\section{Pressing Unsolved Problems Requiring Simulation}
\label{sec:unsolved}



As the preceding chapters have illustrated in many ways, the field of TDEs remains filled with important unsolved problems.   For many, the physics is so complex that analytic methods have little power: although it is possible in many cases to study idealized problems analytically, it is often at the cost of ignoring mechanisms of comparable influence to the one treated, or of imposing artifical symmetries or boundary conditions, or of being restricted to implausible corners of parameter space (most often by being limited to linear perturbation theory).  By contrast, numerical methods can permit the simultaneous consideration of multiple mechanisms, are capable of dealing with highly asymmetric geometries, and make no distinction between small and large amplitude fluctuations.

In this chapter, we will begin by posing a number of TDE problems for which numerical work promises to yield major rewards in understanding.  In the remainder of the chapter, we will critically discuss the simulation methods (i.e., codes) currently available in terms of how they measure up to the demands of these questions, and then apply the same standards to methods currently under development.

\subsection{Magnetic Seeding of Bound Debris}
\label{sec:magnetic_debris}

Once the bound remains of a destroyed star return from the apocenter of their first orbit, they join an accretion flow around the black hole, as discussed in the contribution of Bonnerot et~al. (2019).  In most accretion flows, inward motion is limited by angular momentum transport, and the dominant mechanism providing that transport is correlated magnetic stresses in MHD turbulence driven by the magnetorotational instability (MRI) \citep{BH98}.  In TDE accretion flows, although stream deflection in oblique shocks initially dominates angular momentum transport, these shocks decay over a span of 5--10 orbital periods of the most-bound debris \citep{Shiokawa2015}.  Any further inward flow must be due to some other mechanism, perhaps a version of the correlated MHD turbulence familiar from other accretion problems.

However, little is known about how MHD turbulence might develop in these circumstances.  Presumably, the magnetic field of the star is bequeathed to the bound debris because the matter formerly in the star should stay highly-conductive, consistently supporting flux-freezing, but it is unclear what intensity or geometry it will have once the debris settles into the accretion flow.  There are more questions about the development of the MRI because our entire knowledge-base hitherto has been built upon the assumption of circular orbits, while the debris orbits of a TDE are highly elliptical, with initial eccentricities $e$ within a few percent of unity.  Only in the past year has it been shown that the MRI remains linearly unstable in elliptical disks \citep{Chan2018} and grows with a rate comparable to that found in circular-orbit disks, but to date there has been no work on its nonlinear saturation.

Such knowledge is critical for understanding the longer-term evolution of the accretion flow.   It is also critical for understanding the possible launch of relativistic jets.  The phenomenology of several TDEs points strongly to the existence of relativistic jets \citep{Burrows:2011a,Bloom:2011a,Cenko:2012b,Brown:2015a}, but the magnetic flux required to support a jet of sufficient strength through the Blandford-Znajek mechanism (or some variant) is $>\sim 10^{29} M_6 L_{46}$~G~cm$^2$, considerably greater than the magnetic flux of typical stars, $\sim 10^{22}$~G~cm$^2$.  Thus, field evolution after disruption is also of great interest.

These desiderata point strongly to the need for simulations able to follow the magnetic field from an origin in the star, through its expulsion into the much larger volume around the black hole occupied by the debris, and continuing over the course of whatever subsequent events may amplify (or diminish) its intensity, shape its geometry, or reconfigure its flux.   These simulations will not be easy because they must span a very large dynamic range in lengthscale, from within the star to the span of the debris orbits: the radius of a $1 M_\odot$ main sequence star is $\simeq 0.5 M_6^{-1} r_g$, while the orbital semi-major axes are larger by a factor of at least $(M_{\rm BH}/M_*)^{2/3} \sim 10^4 M_6^{2/3}$ (here $M_6 \equiv M_{\rm BH}/10^6 M_\odot$ and $r_g \equiv GM/c^2$ is the gravitational radius corresponding to the black hole's mass).  Any such simulation must also ensure that the magnetic field is kept divergence-free to very high accuracy in order to avoid unphysical magnetic forces. 

\subsection{Radiation Flow and Forces}
\label{sec:rad_flow}

The only information we have on TDEs (until they can be detected as gravitational wave events, as suggested by \citet{Kobayashi:2004a}) is through photons.  If we're to claim any real understanding of them, we must therefore be able to show how the dynamics we think control them lead to the electromagnetic spectra and lightcurves we observe.  To accomplish this will demand a greate deal of effort because the EM radiation properties span a wide range of wavelengths, vary in time (these are, after all, flare events), and likely depend on viewing-angle (e.g., relative to the star's orbital axis or the black hole's spin axis).   Tracing radiation properties is also a vital part of understanding the dynamics themselves because at many locations and times during a TDE radiation forces can be important.  This is the case within the initial star if its mass is $> \sim 10 M_\odot$, in shocked debris \citep{Shiokawa2015,Jiang:2016a}, and in the accretion flow close to the black hole because accretion rates can easily be in the Eddington range \citep{Rees:1988a,Ulmer:1999a,Krolik:2016a}.  Even when the radiation fluxes are too small to cause significant forces, photons carrying off heat frequently play a significant role in determining the gas's equation of state, and therefore have indirect dynamical influence.   This last role is of special importance to the flow near the black hole because photon trapping may be substantial when the mass accretion rate is comparable to or greater than Eddington.  Thus, for all these reasons, incorporating radiation transport (at least at the level of gray opacity) into the dynamical simulations is another important goal.

There is, however, good reason why as of yet this has only rarely been done---simultaneous solution of the radiation transfer problem and the equations of hydrodynamics can be both complex and computationally very expensive.  Although the diffusion approximation, or extensions of it like flux-limited diffusion (FLD) and the M1 closure, are excellent assumptions where the optical depth to the surface is very large, they can produce spurious effects when the optical depth becomes order unity or less \citep{McK2014}.  For this reason, diffusion-based schemes are not reliable near or outside an external surface, and this is, of course, precisely where the radiation we see is determined.   Moreover, over time, a bad photospheric boundary condition (which is what FLD and M1 provide) can influence the intensity of radiation deep inside the system where locally the diffusion approximation is entirely valid.   Credible results may, in the end, rest upon genuine multi-angle radiation transfer solutions---which may also need to be time-dependent---and possibly their extension to multi-frequency calculations.  Such calculations can be extremely computationally expensive both because of the large number of equations solved and because time-dependent transfer can force time-steps considerably smaller than those required for the fluid motion.

Further complexity and cost come from spectral considerations.  Spectral properties are often determined by the interaction of many individual atomic processes because the total opacity of an astrophysical gas depends on a sum over its many line and edge features, and these in turn depend on the gas's ionization balance and excited populations (which, in regions of line formation, are frequently not in thermodyamic equilibrium).   Still further complexity is introduced because the portion of the flow near the black hole is necessarily relativistic.  Radiation-matter interactions are almost always most simply described in the rest-frame of the matter, but in a relativistic flow, there are significant frame-shifts between each pair of adjacent cells in a simulation.

\subsection{The Influence of Black Hole Spin}
\label{sec:spin}

For main sequence stars suffering tidal disruption, the tidal radius is never more than several tens of gravitational radii.  General relativistic effects can therefore be important, and in more than one way.   One effect whose influence has only begun to be explored is the gravitomagnetic (Lense-Thirring) torque exerted by the black hole on orbiting matter.  If the black hole has any spin at all, and the stellar orbital plane does not lie exactly in the equatorial plane of the Kerr spacetime, gravitomagnetic torque drives a precession of the orbital plane at a rate $2 (a/M) (r/r_g)^{-3/2} \times$ the orbital frequency (here $a/M$ is the black hole spin parameter).  Because it is hard to imagine any reason for the stellar orbit's orientation to be correlated with the black hole's spin direction, this precession should occur generically in TDEs.  Moreover, it should do so throughout the event: while the star passes through its orbital pericenter and begins to fall apart, when the bound debris returns to the pericenter region, and during the evolution of the accretion flow created by debris return.   Much interesting behavior could result from this precession \citep{Guillochon:2015b}.

Ever since the work of \citet{BP75}, it has been generally believed that, granted sufficient time, the inner regions of an accretion flow oriented obliquely relative to its central black hole should become aligned with the black hole's spin.  However, there remains much controversy about how rapidly this takes place, under which circumstances, and by what specific mechanisms \citep{Hatchett1981,PP83,Nelson00,Lodato10,SKH13,Fragile14,HK15,HK18,Tchekh18}.   The issue of evolutionary timescale is especially important for TDEs, which are intrinsically transient events; because they may evolve on timescales that are not a great many orbital timescales, while the precession time can be a large multiple of an orbital period, non-equilibrium is a very real issue in this context.   Thus, even the extensive (and disputatious) literature on this topic, which has generally focused on longer-lived systems, gives only limited guidance.

This is a problem that positively calls out for numerical simulation: it involves nonlinear, general relativistic hydrodynamics in intrinsically 3-d geometry.  The need for simulation is underlined by the fact that this problem is all about angular momentum transport and in accretion disks that means MHD turbulence must be included.  Although capable general relativistic MHD codes exist, it remains a challenging problem to treat because the precession timescale is long compared to the orbital timescale and because very good resolution is necessary to ensure that numerical diffusion on the gridscale does not masquerade as physical angular momentum transport.

\subsection{Self-gravity}
\label{sec:self-gravity}

The star's self-gravity is, of course, essential to determining its structure in isolation as well as the trade-off between this force and the external tidal gravity of the black hole when the star approaches sufficiently close.    Where its effect remains poorly understood is in the debris.  \citet{Kochanek:1994a} pointed out that self-gravity could cause confinement of the debris in the plane perpendicular to the stream's extension; at the same time, however, he also noted that modest injections of entropy, e.g., from H recombination or internal shocks, could substantially counteract its action \citep{Kochanek:1994a}.  More recently, \citet{Coughlin:2016c} argued that if the stream evolution is perfectly adiabatic, whether self-gravity is vital or negligible depends on the value of $d\ln p/d\ln \rho$.  It is also possible that the innermost portion of the debris is subject to significant self-gravity, while a lower-density halo is not \citep{Yalinewich19,Steinberg19}.

The degree to which self-gravitational confinement occurs might be important to later stages of the event because compression of the stream cross-section could affect the deceleration of returning streams when they strike the accretion flow created by earlier-returning debris.   However, the epoch of its possible importance ends once most of the returning mass has suffered at least one shock; at that point, the additional entropy created overwhelms any self-gravity because these shocks have speeds $\sim 10^{3\pm 0.5}$~km~s$^{-1}$, comparable to stellar interior sound speeds, but occur in gas of far lower density.

Once again, the nonlinearity and geometrical complexity of the problem make it intractable to anything but computational methods.    Also once again, there are technical challenges.    Several of them stem from the long, narrow geometry of the debris stream.   This shape automatically implies that self-gravity is important in only a small fraction of the volume around the black hole.   It also means that high spatial resolution is necessary because the stream width is often a very small fraction of its distance from the black hole.  It also raises the importance of local methods.   In the early stages of a disruption, when the material is still close enough to the black hole for relativistic effects to be significant, any global self-gravity calculation would have to be framed in terms of the Einstein Field Equations rather than the Poisson Equation, although it might be possible to treat the Field Equations in a perturbative fashion.  Even in a local solution (e.g., within a few stellar radii of the star's center-of-mass) it is still necessary to be careful about relativistic considerations (for one method, see \citet{Ryu2+2019}).  Later, when the matter is farther away, only the matter within a few stream thicknesses of a point contributes significantly to the local self-gravity.

Other complications arise because of the sensitivity to entropy in the balance between pressure and self-gravity.   In other words, it is necessary to be careful about radiative cooling or phase changes (like atomic recombination).





\section{Prerequisite Capabilities}
\label{sec:prereq}

This survey of major TDE problems for which simulations would be valuable reveals that numerous physics elements beyond classical hydrodynamics are demanded: general relativity; MHD; radiation transport and radiation forces; self-gravity (Newtonian locally, but requiring reconciliation with the relativistic background); non-adiabatic equations of state.   None of the scientific problems posed demands all of these, but all require at least several of them.

Progress on these scientific problems through simulations also requires overcoming a number of technical difficulties: large dynamic range in lengthscales and timescales; incorporating a phase space (photon direction and energy) beyond the configuration space of fluid dynamics; and a physics repertory that varies with position and time.   Particularly the first two difficulties can pose severe practical problems due to their magnification of the computing time needed.

\section{Currently-Available Tools}
\label{sec:current}

We next turn to an evalution of how existing simulation methods measure up the prerequisites just listed.

\subsection{The {\it HARM} Family}
\label{sec:HARM}

A number of codes in current use for tidal disruption problems all descend from {\it HARM-2D} \citep{GMT2003}.  They share a number of characteristics: all are general relativistic, treat 3 spatial dimensions, employ an intrinsically-conservative fluid algorithm using the Lax-Friedrichs approximation to the Riemann problem, and use a constrained transport (CT) algorithm to update the magnetic field and preserve $\nabla \cdot \vec B = 0$.  Because they are relativistic, a fully self-consistent calculation of gas self-gravity would require the solution of the Einstein Field Equations, but the associated computational load is so great that none of them attempts it.

Although sharing a great deal, they differ from one another in several ways.  {\it HARM3D} \citep{NKH2009,Noble2012} offers complete flexibility in coordinate and spacetime definition, including time-dependent spacetimes.  Its fundamental coordinate system is ``index-space" for the spatial arrays, permitting grid-cells whose physical dimensions can vary in any fashion prescribed by the user.  {\it H-AMR} \citep{Liska2018}, on the other hand, is restricted to the Kerr spacetime in Kerr-Schild spherical coordinates.   However, because it runs on GPUs rather than CPUs, its processing speed is at least an order of magnitude greater than {\it HARM3D}.  In addition, it has the capacity for adaptive mesh refinement (AMR) on blocks of cells.  Both {\it HARM3D} and {\it H-AMR} offer comparatively primitive thermodynamics: heat can be lost by radiation only through optically-thin cooling functions ($\nabla_\mu T^{\mu\nu} = - {\cal L}u^\nu$, for stress-energy trensor $T^{\mu\nu}$, 4-velocity $u^\nu$, and cooling function ${\cal L}$).  {\it KORAL} \citep{Sadowski2013,Sadowski2014} shares the spacetime restrictions of {\it H-AMR}, but introduces quasi-diffusive radiation transfer through the ``M1 closure" approximation, as well as the forces exerted by radiation fluxes.  The M1 closure's essential assumption is that the intensity distribution is axisymmetric around the direction of the mean flux.

\subsection{{\it Flash}}
\label{sec:Flash}

{\it Flash} \citep{Fryxell2010} differs from the {\it HARM} family in a number of respects.  It is Newtonian (or special relativistic), but not general relativistic, and it offers a number of Riemann solution choices (Roe, HLL, HLLC, and HLLD in addition to Lax-Friedrichs).  Most significantly, because its primary version is Newtonian, it can compute self-gravity in terms of the Poisson equation (a number of different algorithms are implemented: multigrid, multipole, FFT, and ``tree").  Like {\it KORAL}, it can solve radiation transfer problems in the semi-diffusive approximation and thereby account for radiation forces, but does so via flux-limited diffusion.  Like {\it H-AMR}, it offers AMR.

\subsection{{\it Athena}}
\label{sec:Athena}

{\it Athena} (\citealt{Stone2008}) is a finite-volume grid based code that solves the Newtonian magneto-hydrodynamic (MHD) equations. It uses constrained transport (CT) to evolve the magnetic fields and a Godunov scheme for shock capturing. In addition to the ideal MHD equations, other terms to describe different physical processes such as resistivity, viscosity, ambipolar diffusion, Hall effects and self-gravity can be included. The radiative transfer equation can be solved based on either the Variable Eddington Tensor method (\citealt{Jiang2012,Davis2012}) or an algorithm that solves the time-dependent equation for specific intensities directly (\citealt{Jiang2014}). Coupled with the MHD equations, either radiation module can be used to determine the thermal properties of the gas self-consistently and predict the observed bolometric lightcurve. The original {\it Athena} code has only Cartesian and cylindrical (\citealt{SkinnerOstriker2010}) coordinate systems with a fixed resolution. It is useful for studying the local physics of TDE streams (\citealt{Salbietal2014,Jiangetal2016}), but it would be difficult to use it for a global TDE simulation.

Some of these limitations are removed by the new version of the code {\it Athena++} (Stone et al., in preparation), which solves the same set of equations as the original {\it Athena} code, but also offers Cartesian, cylindrical and spherical polar coordinate systems with adaptive mesh refinement. A constant time step is currently adopted for all refinement levels.  The radiation modules written for the original version of {\it Athena} also work with {\it Athena++} \cite{Jiang2017}.

In addition, {\it Athena++} provides two new physics modules. One supports general relativistic magneto-hydrodynamics.  It is based on advanced Riemann solvers and staggered-mesh constrained transport (\citealt{Whiteetal2016})and can be used to study the effects of general relativity in TDEs (note that the radiation modules are not consistent with relativistic dynamics).  A second (appropriate only when the spacetime is nearly flat) computes self-gravity in the interior of a bounded region in cylindrical coordinates using an eigenfunction expansion for the azimuthal and vertical dependence and a tridiagonal matrix solver for the radial dependence (the radial grid can be either uniform or logarithmic).  To create appropriate boundary conditions for such a finite region, it iterates between the interior solution and an exterior solution that calculates the ``image charge" necessary to make the interior solution reach the appropriate boundary condition at infinity \citep{Moon19}.  
 
\subsection{Moving Mesh Codes}
\label{sec:movemesh}
``Moving-mesh" is an umbrella term for a class of methods that includes both fully unstructured Lagrangian hydrodynamics schemes such as {\em AREPO} \citep{Springel2010} and codes that implement a moving, but structured mesh, that is tailored to a specific problem geometry. All such schemes potentially reduce advection errors as compared to using the same solver on a fixed grid, while the fully general codes additionally offer an adaptive capability that can be coupled to high-order Godunov schemes. Examples include {\em AREPO}, {\em RICH} \citep{yalinewich15} and {\em ChaNGa} \citep{chang17}, all of which are based on a Voronoi tessellation of the simulation domain. 

A unique numerical consideration relevant to moving mesh codes is the presence of ``mesh noise" that originates from the remapping operations needed as the flow evolves. In addition, current codes that implement general moving meshes do not conserve angular momentum to machine precision, and the adequacy of angular momentum conservation would therefore need to be monitored for long term simulations where it is important (e.g., accretion disks).  However, in a situation such as a Keplerian disk where a high degree of symmetry is guaranteed, a moving mesh whose motion is constrained to follow a known, dominant component of the motion has advantages over a fully unstructured one \citep{duffell16} because it reduces these noise sources.
In common with particle-based codes where forces are computed on a pairwise basis, angular momentum is  typically well-conserved globally. Local conservation of specific angular momentum, on the other hand, is dependent on the time integration scheme, and was relatively poor for first generation moving mesh codes. \cite{pakmor16}  discuss one approach to obtaining improved angular momentum conservation within {\it AREPO}.

Existing codes include different subsets of the physics needed for TDE simulations.  When implemented, self-gravity is typically calculated using a tree algorithm \citep[e.g.][]{barnes86}. A constrained transport scheme for MHD is possible \citep{mocz16}, as is a method based instead on the vector potential \citep{Fragile18}.  Most do not include radiation physics or relativity (even in a fixed metric formulation), although these capabilities do not pose intrinsic problems.

Another existing option for TDE studies is {\it GIZMO} \citep{hopkins15}, which offers a choice between SPH and a mesh-less scheme that implements an arbitrary Lagrangian-Eulerian finite-volume method based on Riemann solvers. {\it GIZMO} conserves angular momentum to machine precision, and uses divergence cleaning for MHD.  However, there are indications that both options suffer from excessive numerical dissipation and incomplete divergence-cleaning when treating nonlinear MHD turbulence \citep{hopkins19}. 

These codes have been employed on TDE problems only within the past year \citep{Goicovic19,Yalinewich19,Steinberg19}, even though their key strengths---the ability to model a range of scales adaptively, follow supersonic flows with reduced diffusion, and capture shocks---are in principle well-matched to the problem.

\subsection{{\it SPH}}
\label{sec:SPH}

Smoothed Particle Hydrodynamics (SPH, \citealt{lucy77,gingold77}) is a mesh-less Lagrangian algorithm that has been widely used, especially in cosmological simulations and hydrodynamic simulation of accretion disks and star formation. Its application to simulations of TDEs is especially convenient in view of its Lagrangian approach, given the large dynamic range in lengthscales between the stellar interior and the extended debris flow.  It is also computationally efficient in the sense that most of the computational domain is ``empty'', with the stellar debris occupying only a limited volume.  In fact, some of the earliest applications of the SPH method were in the context of TDEs. \citet{Nolthenius82} used $\sim 40$ SPH particles to simulate the disruption of a $1M_{\odot}$ star by a $10^4M_{\odot}$ black hole, and \citet{Bicknell83} used 500 SPH particles to simulate a highly penetrating encounter of a $1M_{\odot}$ star with a $10^5M_{\odot}$ black hole.  The goal of the latter simulation was to test the possibility of tidal detonation of the star, a fashionable topic at the time. 

In SPH, the fluid is discretized into finite mass elements (called ``particles'') that evolve according to the Euler-Lagrange equations obtained from a variational principle formulation of fluid dynamics \citep{Eckart60}. Thus, the set of SPH particles represent a Hamiltonian system and as such exactly conserve linear and angular momentum, as well as energy (see \citealt{Price12} and \citealt{Springel2010} for recent reviews). The fluid properties at one particle's position are computed by averaging the properties of neighbouring particles that lie within a ``smoothing region'' around it; various weighting functions are used, but in general they decline with distance from the particle.  The typical size of this region is called the ``smoothing length'' $h$. Typically, the smoothing length is chosen such as to have a constant mass (i.e., number of particles) inside the smoothing sphere. Thus, in SPH resolution automatically follows density and in this sense is a naturally ``adaptive'' method; this adaptivity does not, however, automatically recognize local gradient scales.  It is also important to note that the smallest ``resolved" mass is the mass within a smoothing volume, not the mass of an individual particle; moreover, the mass, energy, and momentum of a particular smoothing volume are subject to Poisson fluctuations due to the finite number of particles within that volume (often chosen to be a few dozen).

Particular care should be given to the handling of shocks and discontinuities in SPH \citep{Price08}. In particular, to resolve a shock, SPH typically requires an artificial bulk viscosity. It can be shown \citep{Lodato10} that, in the absence of switches, the artificial viscosity scales $\propto c_{\rm s}h$, where $c_{\rm s}$ is the gas sound speed. Thus, low density regions, which also have large smoothing lengths $h$, can suffer from large artificial viscosity.  In most cases, however, such dissipation can be effectively limited to the shock region by using suitable switches, such as the \citet{Morris97} switch, or the \citet{Cullen10} switch. This can be particularly important for simulations that try to follow the formation of a disc after a TDE, since at the beginning of the fallback phase, the gas density is bound to be low, and great care should be taken in ensuring that the results are not affected by excessive numerical dissipation.

Another interesting aspect to consider is the inclusion of relativistic effects. SPH naturally lends itself to a fully general relativistic implementation (in a fixed metric), as shown by \citet{Monaghan01}, but it has taken a number of years to complete development of this feature.  This effort began with \citet{Hayasaki2013}, who used the modified pseudo-Newtonian potential of \citet{Wegg2012} to include apsidal precession for particles with nearly-parabolic orbits, and \citet{tejeda13}, who introduced a ``generalized Newtonian potential'' that reproduces test-particle motion in a Schwarzschild spacetime very well for orbits that are identically parabolic, but has unspecified errors for orbits with non-zero binding energy, such as the tidal debris.  This treatment of particle motion as well as a first-order post-Newtonian approximation to Lense-Thirring torque have been added to the {\it PHANTOM} code \citep{Price17} and used in TDE simulations \citep{Bonnerot16}; a full second-order pN approximation was developed by \citet{Hayasaki2016}.  Very recently, a fully relativistic version has been published, using a formalism based on entropy conservation rather than energy conservation \citep{Liptai19}.  However, no provision for relativistically-consistent calculation of stellar self-gravity in SPH simulations has yet been made.

An algorithm for computing 3D radiation transfer in the flux-limited diffusion approximation exists \citep{Whitehouse2005}, but has not yet been used on TDE problems.

\section{Tools under Development}
\label{sec:newtools}

We turn next to new tools on the horizon, comparing them as well to the prerequisites for progress on TDE problems.

\subsection{{\it Patchwork}}
\label{sec:patch}

Many physical systems, including many in astrophysics, contain local inhomogeneities where the local geometric symmetry, or physical lengthscales, or relevant physical processes are different from those elsewhere in the system.  These contrasts pose great difficulties to simulations in which the entire problem is assigned to a single program.  Graded meshes, whether determined in advance or adaptively, can help with contrasting lengthscales, but it is very difficult to adjust the equations to be solved or the symmetry of the grid within a single program.

``Multipatch" systems enable a number of independent programs, each governing a particular region, to solve a unified physical problem by exchanging boundary condition data.  A particularly flexible example is the {\it Patchwork} infrastructure \citep{Shiokawa2018}.  With this infrastructure, built to be intrinsically general relativistic, multiple patches, which may be stationary with respect to one another or moving, can have wholly independent grids, coordinate systems, and physical equations.  For consistency, they must be regarded as moving through the same spacetime and must share the same time coordinate.  When boundary condition data are exchanged, they are therefore transformed by coordinate (not Lorentz) transformations; the transformations for scalar, 4-vector, and tensor quantities follow the standard rules. In its published form, {\it Patchwork} was restricted to problems without magnetic fields because interpolation of magnetic field data onto a new grid as part of inter-patch boundary data exchange creates inter-patch monopoles.  Since then, it has been extended to include algorithms that remove these monopoles (Avara et~al., in preparation).

{\it Patchwork} presents great promise for tidal disruption simulations because TDEs are a prime example of strong inhomogeneities within a single physical system.  They exhibit extreme contrasts in lengthscale: the radius of a $1 M_\odot$ main-sequence star is only $\approx 0.5 r_g$ when the black hole mass is $10^6 M_\odot$, while the debris orbits extend as far as $\sim 10^4 r_g$.  They have strong contrasts in grid symmetry: the natural symmetry for orbital motion around a black hole implies the use of polar coordinates with an origin at the center of the black hole, while the natural symmetry for the star's self-gravity argues for an origin at its center-of-mass---and this moves relative to the black hole.  They involve contrasts in relevant physics: self-gravity is, of course, essential to a star and can be significant for the initial evolution of the tidal debris, but elsewhere and later in the development of the event it is irrelevant.  In fact, the challenges of TDEs were one of the prime motivations for development of {\it Patchwork}.

It is also worth noting that multipatch systems can present computational as well as physical advantages.  When the coordinates are chosen with reference to local dynamical symmetries, numerical diffusion can be minimized, and it is often possible to design grids with far fewer total cells while maintaining high resolution where it is needed.  Different patches can also have different time-steps, potentially offering avenues for improved load-balancing.

\subsection{{\it General Relativistic Radiation Athena}}
\label{sec:GRRadAthena}

Extending the current radiative	transfer module for {\it Athena} to be fully compatible with relativistic magneto-hydrodynamics will be a major next step; problems in flat spacetime and curved spacetime will be handled by separate modules. This extension will make it possible to study the dynamics of streams from TDEs as they travel around the black hole with self-consistent thermodynamics and radiation forces.	 Although the first use of such a module will undoubtedly assume gray opacity, increasing computing power would permit use of a more general version incorporating frequency-dependent radiation transport, as well as scattering angular distributions sensitive to background physical properties such as magnetic field direction.

The ability to include radiation transport is greatly enhanced by improvements in MHD computing speed.  The MHD module of {\it Athena++} has already been converted to a GPU version and preliminary testing indicates substantial acceleration: it can update as many as $10^8$ cells per second on a single Nvidia Volta GPU, while maintaining 75\% parallelization efficiency on as many as 250 GPUs.  Although initially written for conventional parallelization, the plan is to write both relativistic radiation modules to be compatible with future conversion to GPUs.

\subsection{{\it Moving Mesh and SPH Codes}}
\label{sec:movingmesh}

The intrinsic potential of moving mesh codes for TDE problems has not been fully 
realized for a number of reasons, including the fact that first-generation moving mesh 
codes typically implemented a more limited set of physical processes than more mature 
methods. Restricted access to some codes, together with an initially small base of experienced 
users, may also have been a factor.

Near-future improvements should remove these limitations. In particular, the core TDE physics of coupling radiation transport to hydrodynamics is expected to be available soon in several moving mesh or mesh-less schemes. {\it GIZMO}, for example, now includes a number of radiation transport methods, including flux-limited diffusion, M1, and Monte Carlo ray-tracing \citep{hopkins_grudic19}. Monte Carlo radiation transport is being implemented in the SPH code PHANTOM, and
there is ongoing work to include similar radiation transport capabilities within {\em ChaNGa}, which is in addition being extended to relativistic computations within a fixed metric.

\subsection{{\it New Fluid Algorithms}}
\label{sec:algos}

Astrophysical simulation codes most commonly implement one of two families of algorithms for solving the fluid equations: either finite-volume methods based on Godunov schemes, or SPH. Finite difference and spectral methods are also in use. Many other algorithms have been developed in the fluid dynamics community, some of which may offer advantages for astrophysical problems including TDEs.  A number of groups, for example, have recently developed astrophysical simulation codes based on discontinuous Galerkin methods \citep[e.g.][]{schaal15,kidder17,anninos17}. Discontinuous Galerkin schemes can provide high order accuracy---approaching that of spectral methods---{\em without} the non-local communication costs that are typically the price that high-order accuracy demands. 

\subsection{{\it Programming Model Advances}}
\label{sec:GPUs}

In addition to improvements in physics capability and numerical algorithms, advances in computer architecture and operating systems also promise greater capability in the near future.

The most mature of these new programming models is the use of GPUs, whose rapid growth in use is due to the fact that GPU  performance has increased at a faster rate than CPU performance in recent years. The GPU version of {\em HARM}, {\em H-AMR}, outperforms its CPU cousin by a factor of between 9 and 14 on the general relativistic MHD disk simulation problem considered by \citet{Liska2018}. In comparing codes and methods it must be remembered that it is the overall compute time needed to obtain a solution of a given fidelity that matters, and that an impressive speed-up on one problem does not necessarily generalize to other problems of interest (for example those where other physical effects dominate the computation). Nonetheless, it is clear that large GPU speed-ups are possible for some problems of interest to the TDE community, and it is likely that codes other than {\em HARM} could benefit from the use of GPU technology.

Whatever algorithm is employed, the goal is to obtain an appropriate balance of absolute code performance (on a small number of cores or GPUs) together with good scaling to large numbers of processing elements. This requires identifying and taking advantage of whatever parallelism the problem offers, while minimizing overheads in communication between processors or in breaking down the work into pieces that can be executed independently. The traditional approach to parallelism in astrophysical simulation codes has been based on {\em data decomposition} using MPI. In this approach, each core executes the same code on a different piece of the data, which is usually sub-divided into spatial domains. The domains can be fixed in advance (static domain decomposition), or adjusted dynamically in cases where the workload is expected to vary as the simulation progresses. Data decomposition often suffices to provide good weak scaling, i.e. ever larger numbers  of processors can be efficiently deployed to solve ever larger instances of the same problem. There are, however, drawbacks. The inherent near-synchronicity of the model does not always make full use of the available communication between processors, leading to less efficient strong scaling. This is a weakness because ideally we would like larger computer systems to be able to solve ``small" problems faster, as well as being able to tackle problems that were previously too large to attempt.

Task-based parallelism is a programming model that can overcome some of the problems of the 
data-based approach. The basic idea is that the overall problem is divided into {\em tasks}, 
which can be quite distinct (for example, computing fluxes of energy and momentum across 
boundaries might be one task, while another might be determining the next time step to take). 
Each task comes with a list of dependencies (other tasks whose output is needed before this 
task can be executed) and conflicts. A scheduler assigns tasks dynamically as the simulation 
proceeds, taking note of the dependencies and with the goal of minimizing the number of idle 
cores at any one time. Adopting this more asynchronous model, in which computation tasks are 
heterogeneous and executed in a dynamically-determined order, has many advantages 
\citep[see, e.g.,][]{kidder17}. It is more robust against failures and can potentially ``hide" 
more of the inevitable communication behind useful computation.

With the exception of GPU computing, the benefit that these and other ``next-generation" methods offer for TDE simulations is largely unquantified. That said, the overall TDE problem can be broken down into several sub-problems (initial disruption, circularization, accretion, generation of emission in different bands, jet formation, etc) that have quite distinct numerical requirements.  It is highly likely that there are large gains to be found from the adoption of new algorithms and parallelism models in at least some of those sub-problems.

%
%

\begin{acknowledgements}
This work was partially supported by: NSF Grant AST-1715032 and Simons Foundation Grant 559794 (JHK); by NSF Grant PHY-1748958 (YJ); and by NASA Grant NNX16AI40G (PJA).
We also thank the International Space Science Institute for hospitality.
\end{acknowledgements}
\bibliographystyle{plainnat}
\bibliography{TDEbook,general}                

\begin{thebibliography}{80}
\providecommand{\natexlab}[1]{#1}
\providecommand{\url}[1]{\texttt{#1}}
\expandafter\ifx\csname urlstyle\endcsname\relax
  \providecommand{\doi}[1]{doi: #1}\else
  \providecommand{\doi}{doi: \begingroup \urlstyle{rm}\Url}\fi

\bibitem[{Anninos} et~al.(2017){Anninos}, {Bryant}, {Fragile}, {Holgado},
  {Lau}, and {Nemergut}]{anninos17}
P.~{Anninos}, C.~{Bryant}, P.~C. {Fragile}, A.~M. {Holgado}, C.~{Lau}, and
  D.~{Nemergut}.
\newblock {CosmosDG: An hp-adaptive Discontinuous Galerkin Code for
  Hyper-resolved Relativistic MHD}.
\newblock \emph{\apjs}, 231:\penalty0 17, August 2017.
\newblock \doi{10.3847/1538-4365/aa7ff5}.

\bibitem[{Balbus} and {Hawley}(1998)]{BH98}
S.~A. {Balbus} and J.~F. {Hawley}.
\newblock {Instability, turbulence, and enhanced transport in accretion disks}.
\newblock \emph{Reviews of Modern Physics}, 70:\penalty0 1--53, January 1998.
\newblock \doi{10.1103/RevModPhys.70.1}.

\bibitem[{Bardeen} and {Petterson}(1975)]{BP75}
J.~M. {Bardeen} and J.~A. {Petterson}.
\newblock {The Lense-Thirring Effect and Accretion Disks around Kerr Black
  Holes}.
\newblock \emph{\apjl}, 195:\penalty0 L65, January 1975.
\newblock \doi{10.1086/181711}.

\bibitem[{Barnes} and {Hut}(1986)]{barnes86}
J.~{Barnes} and P.~{Hut}.
\newblock {A hierarchical O(N log N) force-calculation algorithm}.
\newblock \emph{Nature}, 324:\penalty0 446--449, December 1986.
\newblock \doi{10.1038/324446a0}.

\bibitem[{Bicknell} and {Gingold}(1983)]{Bicknell83}
G.~V. {Bicknell} and R.~A. {Gingold}.
\newblock {On tidal detonation of stars by massive black holes}.
\newblock \emph{\apj}, 273:\penalty0 749--760, October 1983.
\newblock \doi{10.1086/161410}.

\bibitem[Bloom et~al.(2011)Bloom, Giannios, Metzger, Cenko, Perley, Butler,
  Tanvir, Levan, O'Brien, Strubbe, De~Colle, Ramirez-Ruiz, Lee, Nayakshin,
  Quataert, King, Cucchiara, Guillochon, Bower, Fruchter, Morgan, and van~der
  Horst]{Bloom:2011a}
Joshua~S Bloom, Dimitrios Giannios, Brian~D Metzger, S~Bradley Cenko, Daniel~A
  Perley, Nathaniel~R Butler, Nial~R Tanvir, Andrew~J Levan, Paul~T O'Brien,
  Linda~E Strubbe, Fabio De~Colle, Enrico Ramirez-Ruiz, William~H Lee, Sergei
  Nayakshin, Eliot Quataert, Andrew~R King, Antonino Cucchiara, James
  Guillochon, Geoffrey~C Bower, Andrew~S Fruchter, Adam~N Morgan, and
  Alexander~J van~der Horst.
\newblock A possible relativistic jetted outburst from a massive black hole fed
  by a tidally disrupted star.
\newblock \emph{Science}, 333:\penalty0 203, 00 2011.

\bibitem[{Bonnerot} et~al.(2016){Bonnerot}, {Rossi}, {Lodato}, and
  {Price}]{Bonnerot16}
C.~{Bonnerot}, E.~M. {Rossi}, G.~{Lodato}, and D.~J. {Price}.
\newblock {Disc formation from tidal disruptions of stars on eccentric orbits
  by Schwarzschild black holes}.
\newblock \emph{\mnras}, 455:\penalty0 2253--2266, January 2016.
\newblock \doi{10.1093/mnras/stv2411}.

\bibitem[{Brown} et~al.(2015){Brown}, {Levan}, {Stanway}, {Tanvir}, {Cenko},
  {Berger}, {Chornock}, and {Cucchiaria}]{Brown:2015a}
G.~C. {Brown}, A.~J. {Levan}, E.~R. {Stanway}, N.~R. {Tanvir}, S.~B. {Cenko},
  E.~{Berger}, R.~{Chornock}, and A.~{Cucchiaria}.
\newblock {Swift J1112.2-8238: a candidate relativistic tidal disruption
  flare}.
\newblock \emph{\mnras}, 452:\penalty0 4297--4306, October 2015.
\newblock \doi{10.1093/mnras/stv1520}.

\bibitem[Burrows et~al.(2011)Burrows, Kennea, Ghisellini, Mangano, Zhang, Page,
  Eracleous, Romano, Sakamoto, Falcone, Osborne, Campana, Beardmore, Breeveld,
  Chester, Corbet, Covino, Cummings, D'avanzo, D'elia, Esposito, Evans,
  Fugazza, Gelbord, Hiroi, Holland, Huang, Im, Israel, Jeon, Jun, Kawai, Kim,
  Krimm, Marshall, M{\'e}sz{\'a}ros, Negoro, Omodei, Park, Perkins, Sugizaki,
  Sung, Tagliaferri, Troja, Ueda, Urata, Usui, Antonelli, Barthelmy, Cusumano,
  Giommi, Melandri, Perri, Racusin, Sbarufatti, Siegel, and
  Gehrels]{Burrows:2011a}
D~Burrows, J~Kennea, G~Ghisellini, V~Mangano, B~Zhang, K~Page, M~Eracleous,
  P~Romano, T~Sakamoto, A~Falcone, J~Osborne, S~Campana, A~Beardmore,
  A~Breeveld, M~Chester, R~Corbet, S~Covino, J~Cummings, P~D'avanzo, V~D'elia,
  P~Esposito, P~Evans, D~Fugazza, J~Gelbord, K~Hiroi, S~Holland, K~Huang, M~Im,
  G~Israel, Y~Jeon, H~Jun, N~Kawai, J~Kim, H~Krimm, F~Marshall,
  P~M{\'e}sz{\'a}ros, H~Negoro, N~Omodei, W~Park, J~Perkins, M~Sugizaki,
  H~Sung, G~Tagliaferri, E~Troja, Y~Ueda, Y~Urata, R~Usui, L~Antonelli,
  S~Barthelmy, G~Cusumano, P~Giommi, A~Melandri, M~Perri, J~Racusin,
  B~Sbarufatti, M~Siegel, and N~Gehrels.
\newblock Relativistic jet activity from the tidal disruption of a star by a
  massive black hole.
\newblock \emph{Nature}, 476:\penalty0 421--421--424--424, 00 2011.

\bibitem[{Cenko} et~al.(2012){Cenko}, {Krimm}, {Horesh}, {Rau}, {Frail},
  {Kennea}, {Levan}, {Holland}, {Butler}, {Quimby}, {Bloom}, {Filippenko},
  {Gal-Yam}, {Greiner}, {Kulkarni}, {Ofek}, {Olivares E.}, {Schady},
  {Silverman}, {Tanvir}, and {Xu}]{Cenko:2012b}
S.~B. {Cenko}, H.~A. {Krimm}, A.~{Horesh}, A.~{Rau}, D.~A. {Frail}, J.~A.
  {Kennea}, A.~J. {Levan}, S.~T. {Holland}, N.~R. {Butler}, R.~M. {Quimby},
  J.~S. {Bloom}, A.~V. {Filippenko}, A.~{Gal-Yam}, J.~{Greiner}, S.~R.
  {Kulkarni}, E.~O. {Ofek}, F.~{Olivares E.}, P.~{Schady}, J.~M. {Silverman},
  N.~R. {Tanvir}, and D.~{Xu}.
\newblock {Swift J2058.4+0516: Discovery of a Possible Second Relativistic
  Tidal Disruption Flare?}
\newblock \emph{\apj}, 753:\penalty0 77, July 2012.
\newblock \doi{10.1088/0004-637X/753/1/77}.

\bibitem[{Chan} et~al.(2018){Chan}, {Krolik}, and {Piran}]{Chan2018}
C.-H. {Chan}, J.~H. {Krolik}, and T.~{Piran}.
\newblock {Magnetorotational Instability in Eccentric Disks}.
\newblock \emph{\apj}, 856:\penalty0 12, March 2018.
\newblock \doi{10.3847/1538-4357/aab15c}.

\bibitem[{Chang} et~al.(2017){Chang}, {Wadsley}, and {Quinn}]{chang17}
P.~{Chang}, J.~{Wadsley}, and T.~R. {Quinn}.
\newblock {A moving-mesh hydrodynamic solver for ChaNGa}.
\newblock \emph{\mnras}, 471:\penalty0 3577--3589, November 2017.
\newblock \doi{10.1093/mnras/stx1809}.

\bibitem[{Coughlin} et~al.(2016){Coughlin}, {Nixon}, {Begelman}, and
  {Armitage}]{Coughlin:2016c}
E.~R. {Coughlin}, C.~{Nixon}, M.~C. {Begelman}, and P.~J. {Armitage}.
\newblock {On the structure of tidally disrupted stellar debris streams}.
\newblock \emph{\mnras}, 459:\penalty0 3089--3103, July 2016.
\newblock \doi{10.1093/mnras/stw770}.

\bibitem[{Cullen} and {Dehnen}(2010)]{Cullen10}
L.~{Cullen} and W.~{Dehnen}.
\newblock {Inviscid smoothed particle hydrodynamics}.
\newblock \emph{\mnras}, 408:\penalty0 669--683, October 2010.
\newblock \doi{10.1111/j.1365-2966.2010.17158.x}.

\bibitem[{Davis} et~al.(2012){Davis}, {Stone}, and {Jiang}]{Davis2012}
S.~W. {Davis}, J.~M. {Stone}, and Y.-F. {Jiang}.
\newblock {A Radiation Transfer Solver for Athena Using Short Characteristics}.
\newblock \emph{\apjs}, 199:\penalty0 9, March 2012.
\newblock \doi{10.1088/0067-0049/199/1/9}.

\bibitem[{Deng} et~al.(2019){Deng}, {Mayer}, {Latter}, {Hopkins}, and
  {Bai}]{hopkins19}
H.~{Deng}, L.~{Mayer}, H.~{Latter}, P.~F. {Hopkins}, and X.-N. {Bai}.
\newblock {Local simulations of MRI turbulence with meshless methods}.
\newblock \emph{arXiv e-prints}, January 2019.

\bibitem[{Duffell}(2016)]{duffell16}
P.~C. {Duffell}.
\newblock {DISCO: A 3D Moving-mesh Magnetohydrodynamics Code Designed for the
  Study of Astrophysical Disks}.
\newblock \emph{\apjs}, 226:\penalty0 2, September 2016.
\newblock \doi{10.3847/0067-0049/226/1/2}.

\bibitem[{Eckart}(1960)]{Eckart60}
C.~{Eckart}.
\newblock {Variation Principles of Hydrodynamics}.
\newblock \emph{Physics of Fluids}, 3:\penalty0 421--427, May 1960.
\newblock \doi{10.1063/1.1706053}.

\bibitem[{Fragile} et~al.(2018){Fragile}, {Nemergut}, {Shaw}, and
  {Anninos}]{Fragile18}
P.~C. {Fragile}, D.~{Nemergut}, P.~L. {Shaw}, and P.~{Anninos}.
\newblock {Divergence-Free Magnetohydrodynamics on Conformally Moving, Adaptive
  Meshes Using a Vector Potential Method}.
\newblock \emph{arXiv e-prints}, December 2018.

\bibitem[{Fryxell} et~al.(2010){Fryxell}, {Olson}, {Ricker}, {Timmes},
  {Zingale}, {Lamb}, {MacNeice}, {Rosner}, {Truran}, and {Tufo}]{Fryxell2010}
B.~{Fryxell}, K.~{Olson}, P.~{Ricker}, F.~X. {Timmes}, M.~{Zingale}, D.~Q.
  {Lamb}, P.~{MacNeice}, R.~{Rosner}, J.~W. {Truran}, and H.~{Tufo}.
\newblock {FLASH: Adaptive Mesh Hydrodynamics Code for Modeling Astrophysical
  Thermonuclear Flashes}.
\newblock Astrophysics Source Code Library, October 2010.

\bibitem[{Gammie} et~al.(2003){Gammie}, {McKinney}, and {T{\'o}th}]{GMT2003}
C.~F. {Gammie}, J.~C. {McKinney}, and G.~{T{\'o}th}.
\newblock {HARM: A Numerical Scheme for General Relativistic
  Magnetohydrodynamics}.
\newblock \emph{\apj}, 589:\penalty0 444--457, May 2003.
\newblock \doi{10.1086/374594}.

\bibitem[{Gingold} and {Monaghan}(1977)]{gingold77}
R.~A. {Gingold} and J.~J. {Monaghan}.
\newblock {Smoothed particle hydrodynamics - Theory and application to
  non-spherical stars}.
\newblock \emph{\mnras}, 181:\penalty0 375--389, November 1977.
\newblock \doi{10.1093/mnras/181.3.375}.

\bibitem[{Goicovic} et~al.(2019){Goicovic}, {Springel}, {Ohlmann}, and
  {Pakmor}]{Goicovic19}
F.~G. {Goicovic}, V.~{Springel}, S.~T. {Ohlmann}, and R.~{Pakmor}.
\newblock {Hydrodynamical moving-mesh simulations of the tidal disruption of
  stars by supermassive black holes}.
\newblock \emph{arXiv e-prints}, February 2019.

\bibitem[{Guillochon} and {Ramirez-Ruiz}(2015)]{Guillochon:2015b}
J.~{Guillochon} and E.~{Ramirez-Ruiz}.
\newblock {A Dark Year for Tidal Disruption Events}.
\newblock \emph{\apj}, 809:\penalty0 166, August 2015.
\newblock \doi{10.1088/0004-637X/809/2/166}.

\bibitem[{Hatchett} et~al.(1981){Hatchett}, {Begelman}, and
  {Sarazin}]{Hatchett1981}
S.~P. {Hatchett}, M.~C. {Begelman}, and C.~L. {Sarazin}.
\newblock {A new look at the dynamics of twisted accretion disks}.
\newblock \emph{\apj}, 247:\penalty0 677--685, July 1981.
\newblock \doi{10.1086/159079}.

\bibitem[{Hawley} and {Krolik}(2018)]{HK18}
J.~F. {Hawley} and J.~H. {Krolik}.
\newblock {Sound Speed Dependence of Alignment in Accretion Disks Subjected to
  Lense-Thirring Torques}.
\newblock \emph{\apj}, 866:\penalty0 5, October 2018.
\newblock \doi{10.3847/1538-4357/aadf90}.

\bibitem[{Hayasaki} et~al.(2013){Hayasaki}, {Stone}, and {Loeb}]{Hayasaki2013}
Kimitake {Hayasaki}, Nicholas {Stone}, and Abraham {Loeb}.
\newblock {Finite, intense accretion bursts from tidal disruption of stars on
  bound orbits}.
\newblock \emph{\mnras}, 434\penalty0 (2):\penalty0 909--924, Sep 2013.
\newblock \doi{10.1093/mnras/stt871}.

\bibitem[{Hayasaki} et~al.(2016){Hayasaki}, {Stone}, and {Loeb}]{Hayasaki2016}
Kimitake {Hayasaki}, Nicholas {Stone}, and Abraham {Loeb}.
\newblock {Circularization of tidally disrupted stars around spinning
  supermassive black holes}.
\newblock \emph{\mnras}, 461\penalty0 (4):\penalty0 3760--3780, Oct 2016.
\newblock \doi{10.1093/mnras/stw1387}.

\bibitem[{Hopkins}(2015)]{hopkins15}
P.~F. {Hopkins}.
\newblock {A new class of accurate, mesh-free hydrodynamic simulation methods}.
\newblock \emph{\mnras}, 450:\penalty0 53--110, June 2015.
\newblock \doi{10.1093/mnras/stv195}.

\bibitem[{Hopkins} and {Grudi{\'c}}(2019)]{hopkins_grudic19}
Philip~F. {Hopkins} and Michael~Y. {Grudi{\'c}}.
\newblock {Numerical problems in coupling photon momentum (radiation pressure)
  to gas}.
\newblock \emph{\mnras}, 483:\penalty0 4187--4196, March 2019.
\newblock \doi{10.1093/mnras/sty3089}.

\bibitem[{Jiang} et~al.(2012){Jiang}, {Stone}, and {Davis}]{Jiang2012}
Y.-F. {Jiang}, J.~M. {Stone}, and S.~W. {Davis}.
\newblock {A Godunov Method for Multidimensional Radiation Magnetohydrodynamics
  Based on a Variable Eddington Tensor}.
\newblock \emph{\apjs}, 199:\penalty0 14, March 2012.
\newblock \doi{10.1088/0067-0049/199/1/14}.

\bibitem[{Jiang} et~al.(2014){Jiang}, {Stone}, and {Davis}]{Jiang2014}
Y.-F. {Jiang}, J.~M. {Stone}, and S.~W. {Davis}.
\newblock {An Algorithm for Radiation Magnetohydrodynamics Based on Solving the
  Time-dependent Transfer Equation}.
\newblock \emph{\apjs}, 213:\penalty0 7, July 2014.
\newblock \doi{10.1088/0067-0049/213/1/7}.

\bibitem[{Jiang} et~al.(2016{\natexlab{a}}){Jiang}, {Guillochon}, and
  {Loeb}]{Jiang:2016a}
Y.-F. {Jiang}, J.~{Guillochon}, and A.~{Loeb}.
\newblock {Prompt Radiation and Mass Outflows from the Stream-Stream Collisions
  of Tidal Disruption Events}.
\newblock \emph{\apj}, 830:\penalty0 125, October 2016{\natexlab{a}}.
\newblock \doi{10.3847/0004-637X/830/2/125}.

\bibitem[{Jiang} et~al.(2016{\natexlab{b}}){Jiang}, {Guillochon}, and
  {Loeb}]{Jiangetal2016}
Y.-F. {Jiang}, J.~{Guillochon}, and A.~{Loeb}.
\newblock {Prompt Radiation and Mass Outflows from the Stream-Stream Collisions
  of Tidal Disruption Events}.
\newblock \emph{\apj}, 830:\penalty0 125, October 2016{\natexlab{b}}.
\newblock \doi{10.3847/0004-637X/830/2/125}.

\bibitem[{Jiang} et~al.(2017){Jiang}, {Stone}, and {Davis}]{Jiang2017}
Y.-F. {Jiang}, J.~{Stone}, and S.~W. {Davis}.
\newblock {Super-Eddington Accretion Disks around Supermassive black Holes}.
\newblock \emph{ArXiv e-prints}, September 2017.

\bibitem[{Kidder} et~al.(2017){Kidder}, {Field}, {Foucart}, {Schnetter},
  {Teukolsky}, {Bohn}, {Deppe}, {Diener}, {H{\'e}bert}, {Lippuner}, {Miller},
  {Ott}, {Scheel}, and {Vincent}]{kidder17}
L.~E. {Kidder}, S.~E. {Field}, F.~{Foucart}, E.~{Schnetter}, S.~A. {Teukolsky},
  A.~{Bohn}, N.~{Deppe}, P.~{Diener}, F.~{H{\'e}bert}, J.~{Lippuner},
  J.~{Miller}, C.~D. {Ott}, M.~A. {Scheel}, and T.~{Vincent}.
\newblock {SpECTRE: A task-based discontinuous Galerkin code for relativistic
  astrophysics}.
\newblock \emph{Journal of Computational Physics}, 335:\penalty0 84--114, April
  2017.
\newblock \doi{10.1016/j.jcp.2016.12.059}.

\bibitem[{Kobayashi} et~al.(2004){Kobayashi}, {Laguna}, {Phinney}, and
  {M{\'e}sz{\'a}ros}]{Kobayashi:2004a}
S.~{Kobayashi}, P.~{Laguna}, E.~S. {Phinney}, and P.~{M{\'e}sz{\'a}ros}.
\newblock {Gravitational Waves and X-Ray Signals from Stellar Disruption by a
  Massive Black Hole}.
\newblock \emph{\apj}, 615:\penalty0 855--865, November 2004.
\newblock \doi{10.1086/424684}.

\bibitem[Kochanek(1994)]{Kochanek:1994a}
Christopher~S Kochanek.
\newblock The aftermath of tidal disruption: The dynamics of thin gas streams.
\newblock \emph{\apj}, 422:\penalty0 508, 00 1994.

\bibitem[{Krolik} et~al.(2016){Krolik}, {Piran}, {Svirski}, and
  {Cheng}]{Krolik:2016a}
J.~{Krolik}, T.~{Piran}, G.~{Svirski}, and R.~M. {Cheng}.
\newblock {ASASSN-14li: A Model Tidal Disruption Event}.
\newblock \emph{ArXiv e-prints}, February 2016.

\bibitem[{Krolik} and {Hawley}(2015)]{HK15}
J.~H. {Krolik} and J.~F. {Hawley}.
\newblock {A Steady-state Alignment Front in an Accretion Disk Subjected to
  Lense-Thirring Torques}.
\newblock \emph{\apj}, 806:\penalty0 141, June 2015.
\newblock \doi{10.1088/0004-637X/806/1/141}.

\bibitem[{Liptai} and {Price}(2019)]{Liptai19}
D.~{Liptai} and D.~J. {Price}.
\newblock {General relativistic smoothed particle hydrodynamics}.
\newblock \emph{\mnras}, January 2019.
\newblock \doi{10.1093/mnras/stz111}.

\bibitem[{Liska} et~al.(2018{\natexlab{a}}){Liska}, {Hesp}, {Tchekhovskoy},
  {Ingram}, {van der Klis}, and {Markoff}]{Liska2018}
M.~{Liska}, C.~{Hesp}, A.~{Tchekhovskoy}, A.~{Ingram}, M.~{van der Klis}, and
  S.~{Markoff}.
\newblock {Formation of precessing jets by tilted black hole discs in 3D
  general relativistic MHD simulations}.
\newblock \emph{\mnras}, 474:\penalty0 L81--L85, February 2018{\natexlab{a}}.
\newblock \doi{10.1093/mnrasl/slx174}.

\bibitem[{Liska} et~al.(2018{\natexlab{b}}){Liska}, {Tchekhovskoy}, {Ingram},
  and {van der Klis}]{Tchekh18}
M.~{Liska}, A.~{Tchekhovskoy}, A.~{Ingram}, and M.~{van der Klis}.
\newblock {Bardeen-Petterson Alignment, Jets and Magnetic Truncation in GRMHD
  Simulations of Tilted Thin Accretion Discs}.
\newblock \emph{arXiv e-prints}, October 2018{\natexlab{b}}.

\bibitem[{Lodato} and {Price}(2010)]{Lodato10}
G.~{Lodato} and D.~J. {Price}.
\newblock {On the diffusive propagation of warps in thin accretion discs}.
\newblock \emph{\mnras}, 405:\penalty0 1212--1226, June 2010.
\newblock \doi{10.1111/j.1365-2966.2010.16526.x}.

\bibitem[{Lucy}(1977)]{lucy77}
L.~B. {Lucy}.
\newblock {A numerical approach to the testing of the fission hypothesis}.
\newblock \emph{\aj}, 82:\penalty0 1013--1024, December 1977.
\newblock \doi{10.1086/112164}.

\bibitem[{McKinney} et~al.(2014){McKinney}, {Tchekhovskoy}, {Sadowski}, and
  {Narayan}]{McK2014}
J.~C. {McKinney}, A.~{Tchekhovskoy}, A.~{Sadowski}, and R.~{Narayan}.
\newblock {Three-dimensional general relativistic radiation
  magnetohydrodynamical simulation of super-Eddington accretion, using a new
  code HARMRAD with M1 closure}.
\newblock \emph{\mnras}, 441:\penalty0 3177--3208, July 2014.
\newblock \doi{10.1093/mnras/stu762}.

\bibitem[{Mocz} et~al.(2016){Mocz}, {Pakmor}, {Springel}, {Vogelsberger},
  {Marinacci}, and {Hernquist}]{mocz16}
P.~{Mocz}, R.~{Pakmor}, V.~{Springel}, M.~{Vogelsberger}, F.~{Marinacci}, and
  L.~{Hernquist}.
\newblock {A moving mesh unstaggered constrained transport scheme for
  magnetohydrodynamics}.
\newblock \emph{\mnras}, 463:\penalty0 477--488, November 2016.
\newblock \doi{10.1093/mnras/stw2004}.

\bibitem[{Monaghan} and {Price}(2001)]{Monaghan01}
J.~J. {Monaghan} and D.~J. {Price}.
\newblock {Variational principles for relativistic smoothed particle
  hydrodynamics}.
\newblock \emph{\mnras}, 328:\penalty0 381--392, December 2001.
\newblock \doi{10.1046/j.1365-8711.2001.04742.x}.

\bibitem[{Moon} et~al.(2019){Moon}, {Kim}, and {Ostriker}]{Moon19}
S.~{Moon}, W.-T. {Kim}, and E.~C. {Ostriker}.
\newblock {A Fast Poisson Solver of Second-Order Accuracy for Isolated Systems
  in Three-Dimensional Cartesian and Cylindrical Coordinates}.
\newblock \emph{arXiv e-prints}, February 2019.

\bibitem[{Morris} and {Monaghan}(1997)]{Morris97}
J.~P. {Morris} and J.~J. {Monaghan}.
\newblock {A Switch to Reduce SPH Viscosity}.
\newblock \emph{Journal of Computational Physics}, 136:\penalty0 41--50,
  September 1997.
\newblock \doi{10.1006/jcph.1997.5690}.

\bibitem[{Nelson} and {Papaloizou}(2000)]{Nelson00}
R.~P. {Nelson} and J.~C.~B. {Papaloizou}.
\newblock {Hydrodynamic simulations of the Bardeen-Petterson effect}.
\newblock \emph{\mnras}, 315:\penalty0 570--586, July 2000.
\newblock \doi{10.1046/j.1365-8711.2000.03478.x}.

\bibitem[{Noble} et~al.(2009){Noble}, {Krolik}, and {Hawley}]{NKH2009}
S.~C. {Noble}, J.~H. {Krolik}, and J.~F. {Hawley}.
\newblock {Direct Calculation of the Radiative Efficiency of an Accretion Disk
  Around a Black Hole}.
\newblock \emph{\apj}, 692:\penalty0 411--421, February 2009.
\newblock \doi{10.1088/0004-637X/692/1/411}.

\bibitem[{Noble} et~al.(2012){Noble}, {Mundim}, {Nakano}, {Krolik},
  {Campanelli}, {Zlochower}, and {Yunes}]{Noble2012}
S.~C. {Noble}, B.~C. {Mundim}, H.~{Nakano}, J.~H. {Krolik}, M.~{Campanelli},
  Y.~{Zlochower}, and N.~{Yunes}.
\newblock {Circumbinary Magnetohydrodynamic Accretion into Inspiraling Binary
  Black Holes}.
\newblock \emph{\apj}, 755:\penalty0 51, August 2012.
\newblock \doi{10.1088/0004-637X/755/1/51}.

\bibitem[{Nolthenius} and {Katz}(1982)]{Nolthenius82}
R.~A. {Nolthenius} and J.~I. {Katz}.
\newblock {The passage of a star by a massive black hole}.
\newblock \emph{\apj}, 263:\penalty0 377--385, December 1982.
\newblock \doi{10.1086/160511}.

\bibitem[{Pakmor} et~al.(2016){Pakmor}, {Springel}, {Bauer}, {Mocz}, {Munoz},
  {Ohlmann}, {Schaal}, and {Zhu}]{pakmor16}
R.~{Pakmor}, V.~{Springel}, A.~{Bauer}, P.~{Mocz}, D.~J. {Munoz}, S.~T.
  {Ohlmann}, K.~{Schaal}, and C.~{Zhu}.
\newblock {Improving the convergence properties of the moving-mesh code AREPO}.
\newblock \emph{\mnras}, 455:\penalty0 1134--1143, January 2016.
\newblock \doi{10.1093/mnras/stv2380}.

\bibitem[{Papaloizou} and {Pringle}(1983)]{PP83}
J.~C.~B. {Papaloizou} and J.~E. {Pringle}.
\newblock {The time-dependence of non-planar accretion discs}.
\newblock \emph{\mnras}, 202:\penalty0 1181--1194, March 1983.
\newblock \doi{10.1093/mnras/202.4.1181}.

\bibitem[{Price}(2008)]{Price08}
D.~J. {Price}.
\newblock {Modelling discontinuities and Kelvin Helmholtz instabilities in
  SPH}.
\newblock \emph{Journal of Computational Physics}, 227:\penalty0 10040--10057,
  December 2008.
\newblock \doi{10.1016/j.jcp.2008.08.011}.

\bibitem[{Price}(2012)]{Price12}
D.~J. {Price}.
\newblock {Smoothed particle hydrodynamics and magnetohydrodynamics}.
\newblock \emph{Journal of Computational Physics}, 231:\penalty0 759--794,
  February 2012.
\newblock \doi{10.1016/j.jcp.2010.12.011}.

\bibitem[{Price} et~al.(2017){Price}, {Wurster}, {Tricco}, {Nixon}, {Toupin},
  {Pettitt}, {Chan}, {Mentiplay}, {Laibe}, {Glover}, {Dobbs}, {Nealon},
  {Liptai}, {Worpel}, {Bonnerot}, {Dipierro}, {Ballabio}, {Ragusa},
  {Federrath}, {Iaconi}, {Reichardt}, {Forgan}, {Hutchison}, {Constantino},
  {Ayliffe}, {Hirsh}, and {Lodato}]{Price17}
D.~J. {Price}, J.~{Wurster}, T.~S. {Tricco}, C.~{Nixon}, S.~{Toupin},
  A.~{Pettitt}, C.~{Chan}, D.~{Mentiplay}, G.~{Laibe}, S.~{Glover}, C.~{Dobbs},
  R.~{Nealon}, D.~{Liptai}, H.~{Worpel}, C.~{Bonnerot}, G.~{Dipierro},
  G.~{Ballabio}, E.~{Ragusa}, C.~{Federrath}, R.~{Iaconi}, T.~{Reichardt},
  D.~{Forgan}, M.~{Hutchison}, T.~{Constantino}, B.~{Ayliffe}, K.~{Hirsh}, and
  G.~{Lodato}.
\newblock {Phantom: A smoothed particle hydrodynamics and magnetohydrodynamics
  code for astrophysics}.
\newblock \emph{ArXiv e-prints}, February 2017.

\bibitem[Rees(1988)]{Rees:1988a}
Martin~J Rees.
\newblock Tidal disruption of stars by black holes of 10 to the 6th-10 to the
  8th solar masses in nearby galaxies.
\newblock \emph{Nature}, 333:\penalty0 523, 00 1988.
\newblock URL \url{http://adsabs.harvard.edu/abs/1988Natur.333..523R}.

\bibitem[{Ryu} et~al.(2020){Ryu}, {Krolik}, {Piran}, and {Noble}]{Ryu2+2019}
Taeho {Ryu}, Julian {Krolik}, Tsvi {Piran}, and Scott~C. {Noble}.
\newblock {Tidal disruptions of main sequence stars -- II. Simulation
  methodology and stellar mass dependence of the character of full tidal
  disruptions}.
\newblock \emph{arXiv e-prints}, art. arXiv:2001.03502, January 2020.

\bibitem[{Salbi} et~al.(2014){Salbi}, {Matzner}, {Ro}, and
  {Levin}]{Salbietal2014}
P.~{Salbi}, C.~D. {Matzner}, S.~{Ro}, and Y.~{Levin}.
\newblock {Oblique Shock Breakout in Supernovae and Gamma-Ray Bursts. II.
  Numerical Solutions for Non-relativistic Pattern Speeds}.
\newblock \emph{\apj}, 790:\penalty0 71, July 2014.
\newblock \doi{10.1088/0004-637X/790/1/71}.

\bibitem[{S{\c a}dowski} et~al.(2013){S{\c a}dowski}, {Narayan},
  {Tchekhovskoy}, and {Zhu}]{Sadowski2013}
A.~{S{\c a}dowski}, R.~{Narayan}, A.~{Tchekhovskoy}, and Y.~{Zhu}.
\newblock {Semi-implicit scheme for treating radiation under M1 closure in
  general relativistic conservative fluid dynamics codes}.
\newblock \emph{\mnras}, 429:\penalty0 3533--3550, March 2013.
\newblock \doi{10.1093/mnras/sts632}.

\bibitem[{S{\c a}dowski} et~al.(2014){S{\c a}dowski}, {Narayan}, {McKinney},
  and {Tchekhovskoy}]{Sadowski2014}
A.~{S{\c a}dowski}, R.~{Narayan}, J.~C. {McKinney}, and A.~{Tchekhovskoy}.
\newblock {Numerical simulations of super-critical black hole accretion flows
  in general relativity}.
\newblock \emph{\mnras}, 439:\penalty0 503--520, March 2014.
\newblock \doi{10.1093/mnras/stt2479}.

\bibitem[{Schaal} et~al.(2015){Schaal}, {Bauer}, {Chandrashekar}, {Pakmor},
  {Klingenberg}, and {Springel}]{schaal15}
K.~{Schaal}, A.~{Bauer}, P.~{Chandrashekar}, R.~{Pakmor}, C.~{Klingenberg}, and
  V.~{Springel}.
\newblock {Astrophysical hydrodynamics with a high-order discontinuous Galerkin
  scheme and adaptive mesh refinement}.
\newblock \emph{\mnras}, 453:\penalty0 4278--4300, November 2015.
\newblock \doi{10.1093/mnras/stv1859}.

\bibitem[{Shiokawa} et~al.(2015){Shiokawa}, {Krolik}, {Cheng}, {Piran}, and
  {Noble}]{Shiokawa2015}
H.~{Shiokawa}, J.~H. {Krolik}, R.~M. {Cheng}, T.~{Piran}, and S.~C. {Noble}.
\newblock {General Relativistic Hydrodynamic Simulation of Accretion Flow from
  a Stellar Tidal Disruption}.
\newblock \emph{\apj}, 804:\penalty0 85, May 2015.
\newblock \doi{10.1088/0004-637X/804/2/85}.

\bibitem[{Shiokawa} et~al.(2018){Shiokawa}, {Cheng}, {Noble}, and
  {Krolik}]{Shiokawa2018}
H.~{Shiokawa}, R.~M. {Cheng}, S.~C. {Noble}, and J.~H. {Krolik}.
\newblock {PATCHWORK: A Multipatch Infrastructure for
  Multiphysics/Multiscale/Multiframe Fluid Simulations}.
\newblock \emph{\apj}, 861:\penalty0 15, July 2018.
\newblock \doi{10.3847/1538-4357/aac2dd}.

\bibitem[{Skinner} and {Ostriker}(2010)]{SkinnerOstriker2010}
M.~A. {Skinner} and E.~C. {Ostriker}.
\newblock {The Athena Astrophysical Magnetohydrodynamics Code in Cylindrical
  Geometry}.
\newblock \emph{\apjs}, 188:\penalty0 290--311, May 2010.
\newblock \doi{10.1088/0067-0049/188/1/290}.

\bibitem[{Sorathia} et~al.(2013){Sorathia}, {Krolik}, and {Hawley}]{SKH13}
K.~A. {Sorathia}, J.~H. {Krolik}, and J.~F. {Hawley}.
\newblock {Magnetohydrodynamic Simulation of a Disk Subjected to Lense-Thirring
  Precession}.
\newblock \emph{\apj}, 777:\penalty0 21, November 2013.
\newblock \doi{10.1088/0004-637X/777/1/21}.

\bibitem[{Springel}(2010)]{Springel2010}
V.~{Springel}.
\newblock {Smoothed Particle Hydrodynamics in Astrophysics}.
\newblock \emph{\araa}, 48:\penalty0 391--430, September 2010.
\newblock \doi{10.1146/annurev-astro-081309-130914}.

\bibitem[{Steinberg} et~al.(2019){Steinberg}, {Coughlin}, {Stone}, and
  {Metzger}]{Steinberg19}
E.~{Steinberg}, E.~R. {Coughlin}, N.~C. {Stone}, and B.~D. {Metzger}.
\newblock {Thawing the frozen-in approximation: implications for self-gravity
  in deeply plunging tidal disruption events}.
\newblock \emph{arXiv e-prints}, March 2019.

\bibitem[{Stone} et~al.(2008){Stone}, {Gardiner}, {Teuben}, {Hawley}, and
  {Simon}]{Stone2008}
J.~M. {Stone}, T.~A. {Gardiner}, P.~{Teuben}, J.~F. {Hawley}, and J.~B.
  {Simon}.
\newblock {Athena: A New Code for Astrophysical MHD}.
\newblock \emph{\apjs}, 178:\penalty0 137--177, September 2008.
\newblock \doi{10.1086/588755}.

\bibitem[{Tejeda} and {Rosswog}(2013)]{tejeda13}
E.~{Tejeda} and S.~{Rosswog}.
\newblock {An accurate Newtonian description of particle motion around a
  Schwarzschild black hole}.
\newblock \emph{\mnras}, 433:\penalty0 1930--1940, August 2013.
\newblock \doi{10.1093/mnras/stt853}.

\bibitem[Ulmer(1999)]{Ulmer:1999a}
Andrew Ulmer.
\newblock Flares from the tidal disruption of stars by massive black holes.
\newblock \emph{\apj}, 514:\penalty0 180, 00 1999.

\bibitem[{Wegg}(2012)]{Wegg2012}
Christopher {Wegg}.
\newblock {Pseudo-Newtonian Potentials for Nearly Parabolic Orbits}.
\newblock \emph{\apj}, 749\penalty0 (2):\penalty0 183, Apr 2012.
\newblock \doi{10.1088/0004-637X/749/2/183}.

\bibitem[{White} et~al.(2016){White}, {Stone}, and {Gammie}]{Whiteetal2016}
C.~J. {White}, J.~M. {Stone}, and C.~F. {Gammie}.
\newblock {An Extension of the Athena++ Code Framework for GRMHD Based on
  Advanced Riemann Solvers and Staggered-mesh Constrained Transport}.
\newblock \emph{\apjs}, 225:\penalty0 22, August 2016.
\newblock \doi{10.3847/0067-0049/225/2/22}.

\bibitem[{Whitehouse} et~al.(2005){Whitehouse}, {Bate}, and
  {Monaghan}]{Whitehouse2005}
S.~C. {Whitehouse}, M.~R. {Bate}, and J.~J. {Monaghan}.
\newblock {A faster algorithm for smoothed particle hydrodynamics with
  radiative transfer in the flux-limited diffusion approximation}.
\newblock \emph{\mnras}, 364:\penalty0 1367--1377, December 2005.
\newblock \doi{10.1111/j.1365-2966.2005.09683.x}.

\bibitem[{Yalinewich} et~al.(2015){Yalinewich}, {Steinberg}, and
  {Sari}]{yalinewich15}
A.~{Yalinewich}, E.~{Steinberg}, and R.~{Sari}.
\newblock {RICH: Open-source Hydrodynamic Simulation on a Moving Voronoi Mesh}.
\newblock \emph{\apjs}, 216:\penalty0 35, February 2015.
\newblock \doi{10.1088/0067-0049/216/2/35}.

\bibitem[{Yalinewich} et~al.(2019){Yalinewich}, {Steinberg}, {Piran}, and
  {Krolik}]{Yalinewich19}
A.~{Yalinewich}, E.~{Steinberg}, T.~{Piran}, and J.~H. {Krolik}.
\newblock {Radio Emission from the unbound Debris of Tidal Disruption Events}.
\newblock \emph{arXiv e-prints}, March 2019.

\bibitem[{Zhuravlev} et~al.(2014){Zhuravlev}, {Ivanov}, {Fragile}, and {Morales
  Teixeira}]{Fragile14}
V.~V. {Zhuravlev}, P.~B. {Ivanov}, P.~C. {Fragile}, and D.~{Morales Teixeira}.
\newblock {No Evidence for Bardeen-Petterson Alignment in GRMHD Simulations and
  Semi-analytic Models of Moderately Thin, Prograde, Tilted Accretion Disks}.
\newblock \emph{\apj}, 796:\penalty0 104, December 2014.
\newblock \doi{10.1088/0004-637X/796/2/104}.

\end{thebibliography}

%
%



\end{document}